\documentclass[12pt]{iopart}
\usepackage{graphicx}

\begin{document}

\title[Calculation of electron density of periodic systems...]{Calculation of electron density of periodic systems using non-orthogonal
localised orbitals}

\author{Lev Kantorovich  and Oleh Danyliv
\footnote[3]{On leave from Institute for Condensed Matter Physics,
National Academy of Science of Ukraine, Ukraine (e-mail:
oleh.danyliv@kcl.ac.uk)} }

\address{Department of Physics, Kings College London, Strand, London WC2R
2LS, UK}

\begin{abstract}
Methods for calculating an electron density of a periodic crystal
constructed using non-orthogonal localised orbitals are discussed.
We demonstrate that an existing method based on the matrix
expansion of the inverse of the overlap matrix into a power series
can only be used when the orbitals are highly localised (e.g.
ionic systems). In other cases including covalent crystals or
those with an intermediate type of chemical bonding this method
may be either numerically inefficient or fail altogether. Instead,
we suggest an exact and numerically efficient method which can be
used for orbitals of practically arbitrary localisation. Theory is
illustrated by numerical calculations on a model
system.
\end{abstract}


\submitto{\JPCM}

\maketitle

\section{Introduction}

Understanding of the electronic structure of extended systems with
a local perturbation, e.g. point defects in the crystal bulk
\cite{Marshall-book} or adsorption of molecules at crystal
surfaces \cite{Chemisorption-Reactivity-1997} is of fundamental
importance in solid state physics and chemistry. One way of
calculating the electronic structure of the mentioned systems is
based on the usage of so-called cluster methods in which a finite
fragment of an extended system (a quantum cluster) is considered
in detail while the rest of the system is treated at a lower level
of theory
\cite{EMC-1,EMC-2,QM/MM,Sauer-Sierka-2000,Hall-Hinde-Burton-Hillier-2000,Rivail3,Murphy-Philipp-Freisner-2000,Barandiaran-1996,Bredow-1999,Petja-2000,Sulimov2002}.
The main problem of any existing cluster based scheme is in
choosing an appropriate termination of the cluster. Usually, the
quantum cluster is surrounded by point charges \cite{QM/MM},
pseudoatoms (see, e.g. \cite{Sulimov2002}), link atoms
\cite{Rivail3,Sauer-Sierka-2000,Murphy-Philipp-Freisner-2000} or
pseudopotentials
\cite{Abarenkov-Tupitsyn-2001,Abarenkov-Tupitsyn-2001r,Sulimov2002}.
In more sophisticated methods the environment region is described
by an electronic wavefunction which could be either frozen
\cite{Barandiaran-1996,Shidlovskaya-2002} or recalculated
self-consistently with that of the quantum cluster region
\cite{Vreven-Morokuma-2000,Sauer-Sierka-2000,Abarenkov-Bulatov-1997,Mo-Gao-2000,Fornili-Sironi-Raimondi-2003}
(a general theory of cluster embedding which comprises most of the
existing cluster schemes is considered in \cite{EMC-1,EMC-2}).

A rather general cluster method based on overlapping (not
orthogonal) localised orbitals is presently being developed in our
laboratory. Our method which is similar in spirit to some
one-electron methods
\cite{Shidlovskaya-2002,Fornili-Sironi-Raimondi-2003,Mo-Gao-2000}
is based on a construction of strongly localised orbitals which
are designed to represent the true electronic density of the
entire system via a combination of elementary densities associated
in simple cases with atoms, ions and/or bonds; these are called
\emph{regions} \cite{Danyliv-LK-2004}. Our intention is to create
a rather general technique which can be valid for systems of
different chemical character, ranging from purely ionic to
strongly covalent (excluding metals). Therefore, the proper choice
of the localisation technique as well as a general method of
calculating electron density out of strongly localised
non-orthogonal orbitals localised within corresponding regions are
crucial for our method to work for a wide range of systems.

The issue of calculating orbitals localised in appropriate regions
for extreme cases of strongly ionic and covalent crystals has been
considered separately \cite{Danyliv-LK-2004}. It is the main
objective of this paper to discuss methods of calculating the
electron density of periodic systems described via localised
non-orthogonal orbitals.

It should be mentioned that the literature on this topic is quite
scarce which is probably explained by the lack of interest (until
recently) to non-orthogonal (non-canonical) molecular orbitals: in
most techniques used in the solid state community orthogonal Bloch
functions are employed in practical calculations. There are only a
few exceptions (see e.g. \cite{Abarenkov-cluster_method-2003}). If
a set of non-orthogonal orbitals is used, the expression for the
electron density is much more complicated since it contains an
inverse of an infinite overlap matrix constructed out of the
non-orthogonal orbitals of the whole system under consideration
\cite{McWeeny}.

As far as we are aware, there have only been two methods developed
which enable calculation of the electron density of a periodic
system from non-orthogonal orbitals. Both methods are based on a
series expansion of the density: while the first method
\cite{Abarenkov-cluster_method-2003} relies on the so-called
cluster expansion of the density, the second one
\cite{Lowdin-1956,Kunz-ext,Kunz-orig-exp} is based on the power
expansion of the inverse overlap matrix. In this paper we analyse
only the second of the methods in detail since the first one is
very complicated and much more difficult to implement. In section
2 we reexamine the second method from the point of view of the
correct density normalisation. Then, in section 3 we suggest an
alternative technique which does not require any series expansion.
Both methods are compared in section 4 using a very simple model
system. The paper is finished with a short discussion and
conclusions in section 5.

\section{Electron density of a periodic system}

\begin{figure}
\begin{center}\includegraphics[%
  height=6cm,
  keepaspectratio]{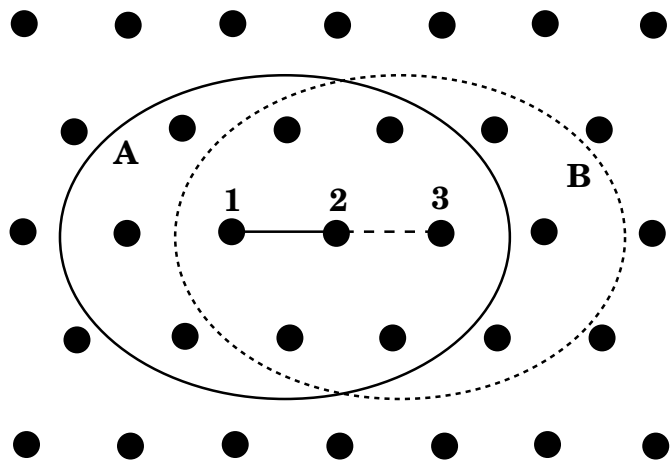}\end{center}

\caption{A schematic of a crystal division into overlapping
regions. Each atom (shown by small black circles) provides a set
of atomic orbitals centred on it. Only two neighbouring regions
$A$ (solid line) and $B$ (dashed line) are shown which physically
represent bonds between atoms 1-2 and 2-3, respectively. All
atomic orbitals centred on atoms within each region contribute to
the localised orbitals associated with this region. It is thus
seen that different regions may have common atomic orbitals if
their borders overlap. In particular, atomic orbitals of atoms 1,
2 and 3 belong to both indicated regions. \label{cap:regions}}
\end{figure}

Let Capital letters $A$, $B$, etc. be used to indicate regions,
while the corresponding small letters $a$, $b$, etc. - localised
orbitals associated with them, i.e. $a\in A$, $b\in B$, etc., see
Fig. \ref{cap:regions}. Each region may have several localised
orbitals. We assume that the orbitals are real. They are expanded
over atomic orbitals centred only on atoms which are inside the
region border. Two localised orbitals belonging to different
regions are not orthogonal either because they have common atomic
orbitals or, if they do not, then due to their exponential tails.

Each region $A$ is prescribed with an even number of $N_{A}$
electrons. Thus, there is a finite number $n_{A}=N_{A}/2$ of
double occupied orbitals associated with the given region $A$. The
localised orbitals $\varphi_{Aa}(\mathbf{r)}$ belonging to the
same region are assumed to be orthonormal; orbitals belonging to
different regions are not orthogonal, i.e the corresponding
overlap integral $S_{Aa,Bb}=\left\langle
\varphi_{Aa}\right.\left|\varphi_{Bb}\right\rangle $ is assumed to
be not zero in general. Note that there might be several regions
within every primitive unit cell in the crystal. Localised
orbitals belonging to physically equivalent regions in different
unit cells are obtained by appropriate lattice translations, i.e.
$\varphi_{Ba}(\mathbf{r)}=\varphi_{Aa}(\mathbf{r-L)}$, where
physically equivalent regions $B$ and $A$ are separated by the
translation vector $\mathbf{L}$.

In general, the spinless electron density takes on the following
general form \begin{equation}
\widetilde{\rho}(\mathbf{r},\mathbf{r^{\prime}})=2\sum_{Aa}\sum_{Bb}\varphi_{Aa}(\mathbf{r)}(\mathbf{S}^{-1})_{Aa,Bb}\varphi_{Bb}(\mathbf{r^{\prime}})\label{eq:density-via-non-othg}\end{equation}
which contains the inverse of the overlap matrix,
$\mathbf{S=\parallel}S_{Aa,Bb}\mathbf{\parallel}$. The overlap
matrix can also be written as a set of finite matrix blocks
$\mathbf{S}_{AB}\mathbf{=\parallel}S_{Aa,Bb}\mathbf{\parallel}$
associated with every pair of regions. Note that for an infinite
crystal the matrix $\mathbf{S}$ has an infinite size. As usual,
the factor of two is due to the fact that each orbital is occupied
by two electrons with opposite spins.

In both summations above localised orbitals from all unit cells
are taken into account. To stress the periodic symmetry of the
crystal, it is useful to rewrite the density in a slightly
different form:\begin{equation}
\widetilde{\rho}(\mathbf{r},\mathbf{r}^{\prime})=\sum_{\mathbf{L}}\rho(\mathbf{r-L},\mathbf{r}^{\prime}-\mathbf{L})\label{eq:Ro-via-translations}\end{equation}
where we introduced a \emph{periodic image} of the density
({}``density image'' for short):

\begin{equation}
\rho(\mathbf{r},\mathbf{r^{\prime}})=2\sum_{Aa}\,^{\prime}\sum_{Bb}\varphi_{Aa}(\mathbf{r)}(\mathbf{S}^{-1})_{Aa,Bb}\varphi_{Bb}(\mathbf{r^{\prime}})\label{eq:image-density}\end{equation}
where in the first sum (indicated by a prime) the summation is run
only over localised orbitals within the single primitive cell
associated with the zero lattice translation; the other summation
runs over all orbitals in the whole infinite system. Note that the
density image is normalised on the number of electrons in the unit
cell only: \begin{equation}
\int\rho(\mathbf{r},\mathbf{r})\textrm{d}\mathbf{r}=\sum_{A}\,^{\prime}N_{A}\label{eq:image-normalisation}\end{equation}

\subsection{Method based on the expansion of the inverse of the overlap matrix\label{sub:S^(-1)_method}}

Following the original prescription by L\"owdin
\cite{Lowdin-1956}, we present the overlap matrix as
$\mathbf{S=1}+\mathbf{\mathbf{\Delta}}$, where the matrix
$\mathbf{\Delta=\parallel}\mathbf{\Delta}_{Aa,Bb}\parallel$ is the
same as the original overlap matrix except for its elements when
$A=B$ which are all equal to zero,
$\mathbf{\Delta}_{Aa,Aa^{\prime}}=0$. Then, one can formally write
a matrix expansion: \begin{equation}
\mathbf{S}^{-1}\mathbf{=(1}+\mathbf{\Delta})^{-1}=\mathbf{1}-\mathbf{\Delta}+\mathbf{\Delta}^{2}-\mathbf{\Delta}^{3}+\ldots\label{eq:expansion-for-S^(-1)}\end{equation}
One can show (using diagonalisation of the matrix $\mathbf{S}$ or
its expansion over the eigenstates) that the expansion
(\ref{eq:expansion-for-S^(-1)}) can only be used if absolute
values of \emph{all} eigenvalues of the matrix $\mathbf{\Delta}$
are less than unity.

Using expansion of Eq. (\ref{eq:expansion-for-S^(-1)}), one
obtains the following expansion for the image density
(\ref{eq:image-density}):\begin{equation}
\rho(\mathbf{r},\mathbf{r^{\prime}})=\sum_{n=0}^{\infty}\rho^{(n)}(\mathbf{r},\mathbf{r^{\prime}})=\sum_{n=0}^{\infty}(-1)^{n}\left[2\sum_{Aa}\,^{\prime}\sum_{Bb}\varphi_{Aa}(\mathbf{r)}(\mathbf{\Delta}^{n})_{Aa,Bb}\varphi_{Bb}(\mathbf{r^{\prime}})\right]\label{eq:image-density-expansion}\end{equation}
Note that a general $n$-th order term (for $n\geq2$) contains
additional $n-1$ summations over all regions due to matrix
multiplications in $\mathbf{\Delta}^{n}$.

In principle, formulae (\ref{eq:Ro-via-translations}) and
(\ref{eq:image-density-expansion}) allow an approximate
calculation of the electron density by terminating the infinite
expansion. Care should be taken, however, in doing so in order to
preserve the correct normalisation of the density.

The zero order contribution, \begin{equation}
\rho^{(0)}(\mathbf{r},\mathbf{r^{\prime}})=2\sum_{Aa}\,^{\prime}\varphi_{Aa}(\mathbf{r)}\varphi_{Aa}(\mathbf{r^{\prime}})\label{eq:0-order-density}\end{equation}
does not contain overlap integrals at all and is normalised to the
total number of electrons in the unit cell. Therefore, if any
higher order terms are kept in the terminated expansion
(\ref{eq:image-density-expansion}), they should be integrated
(normalised) to zero. Consider this point in more detail.

The first order contribution to the image density,\begin{equation}
\rho^{(1)}(\mathbf{r},\mathbf{r^{\prime}})=-2\sum_{Aa}\,^{\prime}\sum_{Bb}\varphi_{Aa}(\mathbf{r)}\Delta_{Aa,Bb}\varphi_{Bb}(\mathbf{r^{\prime}})\label{eq:1-order-density}\end{equation}
 contains different regions $A\neq B$ and thus its contribution to
the charge (or normalisation) becomes: \begin{equation} \Delta
N^{(1)}=\int\rho^{(1)}(\mathbf{r},\mathbf{r})\textrm{d}\mathbf{r}=-2\sum_{Aa}\,^{\prime}\sum_{Bb}\Delta_{Bb,Aa}\Delta_{Aa,Bb}=-2\sum_{A}\,^{\prime}\textrm{Tr}_{A}\left(\mathbf{\Delta}^{2}\right)\label{eq:charge-1}\end{equation}
where the trace $\textrm{Tr}_{A}(\ldots)$ here is calculated with
respect to the localised orbitals belonging to region $A$ only. We
see that the first order term has a finite nonzero charge (in
fact, it is negative).

Any higher order contributions in Eq.
(\ref{eq:image-density-expansion}) for $n\geq2$ contains
additional summations over regions so that equal regions $A=B$ in
the double summation there are also possible. Therefore, every
such contribution, $\rho^{(n)}(\mathbf{r},\mathbf{r^{\prime}})$,
will be split into two terms: a diagonal term, \begin{equation}
\rho_{d}^{(n)}(\mathbf{r},\mathbf{r}^{\prime})=2(-1)^{n}\sum_{Aa,a^{\prime}}\,^{\prime}\varphi_{Aa}(\mathbf{r)}(\mathbf{\Delta}^{n})_{Aa,Aa^{\prime}}\varphi_{Aa^{\prime}}(\mathbf{r^{\prime}})\label{eq:n-order-diag}\end{equation}
in which $A=B$, and a non-diagonal term, \begin{equation}
\rho_{nd}^{(n)}(\mathbf{r},\mathbf{r}^{\prime})=2(-1)^{n}\sum_{Aa}\,^{\prime}\sum_{B(\neq
A),b}\varphi_{Aa}(\mathbf{r)}(\mathbf{\Delta}^{n})_{Aa,Bb}\varphi_{Bb}(\mathbf{r^{\prime}})\label{eq:n-order-nondiag}\end{equation}
associated with $A\neq B$ in Eq.
(\ref{eq:image-density-expansion}). Correspondingly, we obtain the
following contributions to the charge:
\begin{equation}
\Delta
N_{d}^{(n)}=2(-1)^{n}\sum_{Aaa^{\prime}}\,^{\prime}(\mathbf{\Delta}^{n})_{Aa,Aa^{\prime}}S_{Aa,Aa^{\prime}}=2(-1)^{n}\sum_{A}\,^{\prime}\textrm{Tr}_{A}\left(\mathbf{\Delta}^{n}\right)\label{eq:n-order-diag-charge}\end{equation}
\begin{equation}
\Delta N_{nd}^{(n)}=2(-1)^{n}\sum_{Aa}\,^{\prime}\sum_{B(\neq
A),b}(\mathbf{\Delta}^{n})_{Aa,Bb}\Delta_{Bb,Aa}=2(-1)^{n}\sum_{A}\,^{\prime}\textrm{Tr}_{A}\left(\mathbf{\Delta}^{n+1}\right)\label{eq:n-order-nond-charge}\end{equation}

Thus, we see that in any order $n\geq2$ we have $\Delta
N_{nd}^{(n)}\equiv-\Delta N_{d}^{(n+1)}$. This means that the
non-diagonal contribution to the density
(\ref{eq:n-order-nondiag}) is compensated exactly by the diagonal
one (\ref{eq:n-order-diag}) of the next order. For instance, the
non-zero charge (\ref{eq:charge-1}) is to be exactly eliminated by
a charge due to the diagonal second order density; in turn, a
nonzero charge due to non-diagonal second order density is
compensated exactly by the diagonal third order density
contribution, and so on.

This result is very useful since it allows one to balance properly
a terminated expansion for the image density so that it would
correspond (in any order!) to the correct total charge. To do
this, the final expression for the density of any $n$-th order
should also include the diagonal ($A=B$) term from the
contribution of the next order. We stress that this fact was
ignored in the previous applications of this method
\cite{Kunz-ext,Kunz-orig-exp}. We obtain, that the correct $n-$th
order expansion for the image density in the notations of Eqs.
(\ref{eq:n-order-diag}), (\ref{eq:n-order-nondiag}) should have
the form: \begin{equation}
\rho(\mathbf{r},\mathbf{r^{\prime}})\simeq\rho^{[n]}(\mathbf{r},\mathbf{r^{\prime}})\equiv\sum_{i=0}^{n}\left[\rho_{d}^{(i)}(\mathbf{r},\mathbf{r^{\prime}})+\rho_{nd}^{(i)}(\mathbf{r},\mathbf{r^{\prime}})\right]+\rho_{d}^{(n+1)}(\mathbf{r},\mathbf{r^{\prime}})\label{eq:final-n-order-density}\end{equation}
By employing this termination of the series, the normalisation
condition (\ref{eq:image-normalisation}) is satisfied exactly.

Thus, in order to calculate the density up to the $n-$th order,
one has to calculate the matrix elements
$\left(\mathbf{\Delta}^{k}\right)_{Aa,Bb}$ of the matrix $\Delta$
for all powers $k=1,\ldots,n$; in addition, one also need diagonal
$A=B$ elements of $\mathbf{\Delta}^{n+1}$. Then, the contributions
from all density images corresponding to all lattice translations,
Eq. (\ref{eq:Ro-via-translations}), are to be added together to
get the final electron density.

The method described here relies on the convergence of the density
expansion (\ref{eq:image-density-expansion}). The better
localisation of the orbitals $\varphi_{Aa}(\mathbf{r)}$, the
faster convergence and thus smaller number of terms is needed. We
shall demonstrate in section \ref{sec:Results} that in some cases
of not very well localised orbitals one has to consider the
density expansion up to a very high order which makes the
calculation extremely time-consuming. Moreover, if the orbitals
localisation becomes worse than a certain criteria (to be also
discussed in section \ref{sec:Results}), then this method fails
altogether as the expansion diverges. A general and an extremely
efficient technique which is not based on a perturbative expansion
of any kind and can be used for localised orbitals of practically
any degree of localisation is suggested in the next subsection.

\subsection{Method based on the Fourier transform of localised orbitals\label{sub:S(k)-method}}

In Eq. (\ref{eq:density-via-non-othg}) for the electron density,
regions $A$ and $B$ are to be chosen from all unit cells of the
infinite periodic system. It is convenient in this section to
identify explicitly the lattice vector for every localised orbital
in its index. Therefore, in the following we shall use letters
$A$, $B$, etc. only for regions within the zeroth unit cell; in
particular, the orbital $\varphi_{Aa}(\mathbf{r)}$ is assumed to
be from the zeroth cell. Localised orbitals from other cells are
characterised by the combined index $(\mathbf{L}Aa)$, i.e.
$\varphi_{\mathbf{L}Aa}(\mathbf{r)}=\varphi_{Aa}(\mathbf{r-L)}$ is
the $a$-th localised orbital from region $A$ in the unit cell
separated from the zeroth cell by the lattice translation
$\mathbf{L}$.

Correspondingly, Eq. (\ref{eq:density-via-non-othg}) is rewritten
in the following way: \begin{equation}
\widetilde{\rho}(\mathbf{r},\mathbf{r^{\prime}})=2\sum_{\mathbf{L}Aa}\sum_{\mathbf{M}Bb}\varphi_{Aa}(\mathbf{r-L)}(\mathbf{S}^{-1})_{\mathbf{L}Aa,\mathbf{M}Bb}\varphi_{Bb}(\mathbf{r^{\prime}}-\mathbf{M})\label{eq:density-via-transl}\end{equation}
where $\mathbf{L}$ and $\mathbf{M}$ are two lattice translations.
A further transformation is possible here since the overlap
integral $S_{\mathbf{L}Aa,\mathbf{M}Bb}$ depends in fact only on
the difference $\mathbf{M}-\mathbf{L}$ of the lattice
translations. This allows expansion of the overlap integral into
the Fourier integral

\begin{equation}
S_{\mathbf{L}Aa,\mathbf{M}Bb}=\frac{1}{N}\sum_{\mathbf{k}}S_{Aa,Bb}(\mathbf{k})e^{-\textrm{i}\mathbf{k}(\mathbf{L}-\mathbf{M})}\label{eq:direct-overl-via-k-sum}\end{equation}
where the summation is performed over $N$ points $\mathbf{k}$ in
the first Brillouin zone (BZ) and \begin{equation}
S_{Aa,Bb}(\mathbf{k})=\sum_{\mathbf{L}}S_{\mathbf{0}Aa,\mathbf{L}Bb}e^{\textrm{i}\mathbf{kL}}\label{eq:k-overlap-via-direct-sum}\end{equation}
is the corresponding Fourier image. The direct lattice summation
in the last formula is easily terminated due to (usually)
exponential decay of the overlap integrals between localised
orbitals.

Using the Fourier representation of the overlap matrix, one can
exactly calculate its inverse as follows: \begin{equation}
\left(\mathbf{S}^{-1}\right)_{\mathbf{L}Aa,\mathbf{M}Bb}=\frac{1}{N}\sum_{\mathbf{k}}\left[\mathbf{S}^{-1}(\mathbf{k})\right]_{Aa,Bb}e^{-\textrm{i}\mathbf{k}(\mathbf{L}-\mathbf{M})}\label{eq:inv-overl-via-k-sum}\end{equation}
Note that the matrix $\mathbf{S}(\mathbf{k})=\parallel
S_{Aa,Bb}(\mathbf{k})\parallel$ has a finite size of the number of
localised orbitals per unit cell. Therefore, in order to calculate
the inverse of the overlap matrix in direct space, one has to
perform the calculation of $\mathbf{S}^{-1}(\mathbf{k})$ for
finite size matrices for every $\mathbf{k}$ point necessary to
sample the BZ. Substituting Eq. (\ref{eq:inv-overl-via-k-sum})
into Eq. (\ref{eq:density-via-transl}), we arrive at the following
final expression for the electron density: \begin{equation}
\widetilde{\rho}(\mathbf{r},\mathbf{r^{\prime}})=\frac{2}{N}\sum_{\mathbf{k}}\left\{
\sum_{Aa}\sum_{Bb}\varphi_{Aa}(\mathbf{r},\mathbf{k)}\left[\mathbf{S}^{-1}(\mathbf{k})\right]_{Aa,Bb}\varphi_{Bb}^{*}(\mathbf{r^{\prime}},\mathbf{k})\right\}
\label{eq:final-density}\end{equation} where \begin{equation}
\varphi_{Aa}(\mathbf{r},\mathbf{k)}=\sum_{\mathbf{L}}\varphi_{Aa}(\mathbf{r-L)}e^{-\textrm{i}\mathbf{kL}}\label{eq:orbital-Fourier-exp}\end{equation}
is the Fourier expansion of the localised orbital. Due to
exponential decay of the localised orbitals, the summation over
lattice vectors $\mathbf{L}$ in the last expression is in fact
finite.

The obtained formula for the density is exact. In particular, it
contains the periodicity of the lattice built in. It is also
extremely convenient for numerical implementation. Indeed, what is
needed is the calculation of the Fourier images, according to Eq.
(\ref{eq:orbital-Fourier-exp}), of every localised orbital in the
primitive unit cell for every $\mathbf{k}$ point. The summations
in the curly brackets in Eq. (\ref{eq:final-density}) are finite
(limited to the orbitals within the zeroth cell only) and are thus
easily performed. The extend to which the orbitals
$\varphi_{Aa}(\mathbf{r)}$ are localised is reflected by the
number of cells to be taken into account while performing the
lattice summations in Eqs. (\ref{eq:k-overlap-via-direct-sum}) and
(\ref{eq:orbital-Fourier-exp}). Even for orbitals which are not
very well localised, the amount of work needed to perform these
lattice summations is incomparable with the cost of the first
method (section \ref{sub:S^(-1)_method}) which requires including
more terms in the perturbation expansion if the localisation is
not good enough.

\section{Results\label{sec:Results}}

Atomic units are used throughout this section. The application of
the two methods considered in the previous sections is illustrated
here on a simple cubic lattice model containing a single region in
every unit cell. The lattice constant $a$ will be assumed to be
equal to 1 a.u. for simplicity. Each region is represented by a
single localised orbital in a form of a normalised $s$ type
Gaussian

\begin{equation}
\varphi_{\mathbf{L}Aa}(\mathbf{r)\rightarrow\varphi_{L}(r)\equiv\varphi(r}-\mathbf{L)},\,\,\,\varphi(r)=\left(\frac{2\alpha}{\pi}\right)^{3/4}e^{-\alpha\mathbf{r}^{2}}.\label{eq:psi_loc}\end{equation}
By choosing various values for the exponent $\alpha$, one can vary
the degree of localisation of the orbitals. Indeed, the size of
the \emph{spatial extent} of the orbital can be measured in terms
of $r_{eff}=\sqrt{\frac{\ln10}{\alpha}}\simeq1.52\alpha^{-1/2}$,
which corresponds to $e^{-\alpha r_{eff}^{2}}=$0.1. We found this
approach more convenient in our particular case than the
application of the existing localisation criteria (see, e.g.
\cite{Marzari97,Pipek-Mezey}).

For this model system it is possible to do some preliminary
analytical estimates of the convergence of the series
(\ref{eq:expansion-for-S^(-1)}). We know from section
\ref{sub:S^(-1)_method} that the series will converge if
\emph{all} eigenvalues $\Delta_{\lambda}$ of the matrix
$\mathbf{\Delta}=\mathbf{S-1}$ are between -1 and 1. It is easy to
notice that the eigenvalues are in fact given by the Fourier
transforms $\Delta_{\mathbf{k}}$ of the matrix $\mathbf{\Delta}$
which is introduced much in the same way as
$\mathbf{S}(\mathbf{k})$ in Eq.
(\ref{eq:k-overlap-via-direct-sum}). Indeed, because
$\Delta_{\mathbf{L,M}}=\Delta_{\mathbf{0,M-L}}$, one can write:

\begin{equation}
\sum_{\mathbf{M}}\Delta_{\mathbf{L,M}}e^{\textrm{i}\mathbf{kM}}=\left(\sum_{\mathbf{M}}\Delta_{\mathbf{0,M-L}}e^{\textrm{i}\mathbf{k(M-L)}}\right)e^{\textrm{i}\mathbf{kL}}=\Delta_{\mathbf{k}}e^{\textrm{i}\mathbf{kL}}\label{eq:eigenproblem_for_delta}\end{equation}
 This is nothing but the eigenproblem for the matrix $\mathbf{\Delta}$
with $\Delta_{\mathbf{k}}$ being its eigenvalues (numbered by
vectors $\mathbf{k}$ from the BZ) and $\left\Vert
e^{\textrm{i}\mathbf{kL}}\right\Vert $ - eigenvectors. Therefore,
the convergence criteria for the series
(\ref{eq:expansion-for-S^(-1)}) reduces to the inequalities
$\left|\Delta_{\mathbf{k}}\right|<1$ which should be valid for
\emph{any} $\mathbf{k}$. Taking into account the overlap only
between nearest neighbours, we obtain:

\[
\left|\Delta_{\mathbf{k}}\right|=\left|2\delta(\cos(k_{x}a)+\cos(k_{y}a)+\cos(k_{z}a))\right|\le6\delta<1\]
with the overlap between neighbouring orbitals being
$\delta=e^{-\alpha a^{2}/2}$. This results in the following
criterion for the convergence of the L\"owdin expansion (for $a=$1
a.u.):

\begin{equation}
\alpha\succeq\alpha_{1}^{*}=2\mathrm{ln}(6)\approx3.6\label{eq:alpha1}\end{equation}
 Similar analysis which takes into account the next nearest neighbours
gives a very similar estimate of $\alpha_{1}^{*}\approx4.05$.
These estimates correspond to the maximum spatial extent of the
orbitals (\ref{eq:psi_loc}) of the order of $r_{eff}\simeq$0.76
a.u., i.e. there is very small overlap between neighbouring
orbitals which, we recall, are separated by 1 a.u. in the lattice.

The other method based on the Fourier transform of the orbitals
has also its limits which are hidden in the formulae
(\ref{eq:k-overlap-via-direct-sum}) and
(\ref{eq:orbital-Fourier-exp}): if a certain cut-off
$\left|\mathbf{L}\right|\leq r_{c}$ for the direct lattice
summation $\mathbf{L}$ is assumed in the calculation of
$S_{Aa,Bb}(\mathbf{k})$ and $\varphi_{Aa}(\mathbf{r},\mathbf{k)}$,
then there will be some limitations on the allowed degree of
localisation of the orbitals. The required criterion can be worked
out e.g. by analysing the Fourier transform
(\ref{eq:orbital-Fourier-exp}) of the orbital at its maximum in
the centre of the BZ (i.e. of
$\varphi(\mathbf{r}=\mathbf{0},\mathbf{k=0})$) as follows:\[
\sum_{|\mathbf{L}|>r_{c}}\varphi(\mathbf{L)}\ll\sum_{|\mathbf{L}|<r_{c}}\varphi(\mathbf{L)}\]
Replacing the sums by the corresponding volume integrals, we
obtain the following criterion: \[
xe^{-x^{2}}+\frac{\sqrt{\pi}}{2}\textrm{erfc}(x)\ll\frac{\sqrt{\pi}}{4}\]
where $x=r_{c}\sqrt{\alpha}$. The inequality above is satisfied if
$x\succeq2$, i.e. $\alpha\succeq4/r_{c}^{2}$. Assuming that
$r_{c}$ is equal to 4$\div$5 lattice constants, we obtain the
necessary condition for the exponent of the localised orbitals,

\begin{equation}
\alpha\gg\alpha_{2}^{*}\sim0.2\label{eq:alpha2}\end{equation} for
which our Fourier transform method should work. The obtained
critical value of $\alpha_{2}^{*}$ results in the maximum spatial
extent of the orbitals of the order of $r_{eff}\simeq$3.4 a.u.
which corresponds to very diffuse orbitals spreading over more
than six unit cells.

Similar criteria is obtained for the ovelrap integrals as well.
Thus, the method we suggest should have a much wider range of
applicability than the L\"owdin method as far as the degree of
localisation of the non-orthogonal orbitals is concerned since
$\alpha_{1}^{*}\gg\alpha_{2}^{*}$. This conclusion is also
supported by our numerical calculations which we now describe.

Numerical calculations of the necessary powers of the
$\mathbf{\Delta}$ matrix needed for the L\"owdin method were done
in the following way. Since the density is calculated in the same
point $\mathbf{r=}\mathbf{r}^{\prime}$ in Eqs.
(\ref{eq:n-order-diag}), (\ref{eq:n-order-nondiag}) and
(\ref{eq:final-n-order-density}), the regions $A$ and $B$ in these
equations are either the same or not far away from each other.
Therefore, to calculate $\left(\mathbf{\Delta}^{n}\right)_{Aa,Bb}$
one can simply choose a sufficiently big finite cluster of atoms
(in fact, the cluster radius should be at least of the order of
$\frac{n}{2}r_{c}^{*}$, where $r_{c}^{*}$ is the decay length of
the overlap integral) with regions $A$ and $B$ somewhere in its
centre and then calculate the complete ovelrap matrix for it,
$\widetilde{\mathbf{\Delta}}$. Then, by performing the necessary
$n-1$ matrix multiplications, one can calculate
$\left(\mathbf{\Delta}^{n}\right)_{Aa,Bb}$ as
$\left(\widetilde{\mathbf{\Delta}}^{n}\right)_{Aa,Bb}$.

When using the Fourier transform method, we employed the
Monkhorst-Pack (MP) method \cite{Monkhorst-Pack} for the
$\mathbf{k}$ point sampling and the same cut-off distance for the
direct lattice summations in Eqs.
(\ref{eq:k-overlap-via-direct-sum}) and
(\ref{eq:orbital-Fourier-exp}) as in the previous method. In all
our calculations we used the 4x4x4 MP set which was found to be
sufficient in all cases.

\begin{figure}
\begin{center}\includegraphics[%
  height=8cm,
  keepaspectratio]{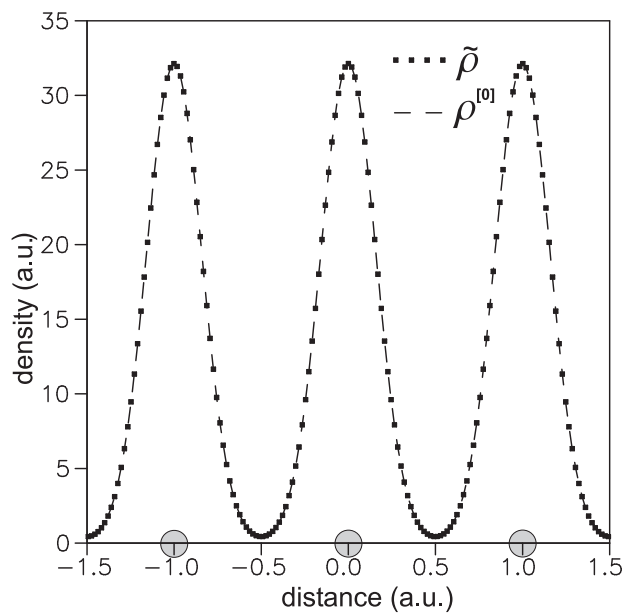}\end{center}

\caption{The exact electron density $\tilde{\rho}(r,r)$, Eq.
(\ref{eq:final-density}), and that based on the zero order
approximation $\rho^{(0)}(r,r)$, Eq. (\ref{eq:0-order-density}),
both calculated along the (100) direction using $\alpha=$10 a.u.
Note that the densities are nearly zero between the localisation
centres shown by grey circles. \label{fig:alpha_10}}
\end{figure}

Results of our calculations for a large value of the exponent
($\alpha\gg\alpha_{1}^{*}\gg\alpha_{2}^{*}$) are shown in
Fig.\ref{fig:alpha_10}. This case corresponds to strongly
localised orbitals as is the case in ionic systems such as MgO and
NaCl. Overlap between orbitals is negligible and even zero order
approximation L\"owdin method, Eq. (\ref{eq:0-order-density}), was
found sufficient to give the correct density. Density curves for
both methods are indistinguishable from each other.

\begin{figure}
\begin{center}\includegraphics[%
  height=8cm,
  keepaspectratio]{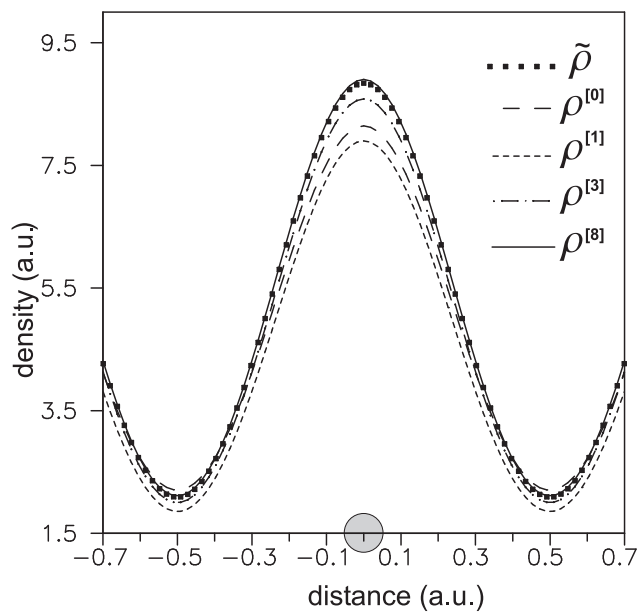}\end{center}

\caption{The exact electron density $\tilde{\rho}(r,r)$ , Eq.
(\ref{eq:final-density}), and the several approximations to it
using Eq. (\ref{eq:final-n-order-density}) with $n=$0, 1, 3 and 8,
all calculated along the (100) direction using $\alpha=$4 a.u.
Note that the density is small (but nonzero) between the
localisation centres. \label{fig:alpha_4}}
\end{figure}

The calculated densities in the intermediate case
($\alpha\sim\alpha_{1}^{*}$) are shown in Fig. \ref{fig:alpha_4}.
This value of $\alpha$ may correspond to ion-covalent and covalent
systems. One can see that high order approximations (up to $n=$8)
of the L\"owdin method, Eq. (\ref{eq:final-n-order-density}), are
needed here to converge the density and thus the calculation is
quite time consuming.

\begin{figure}
\begin{center}\includegraphics[%
  height=8cm,
  keepaspectratio]{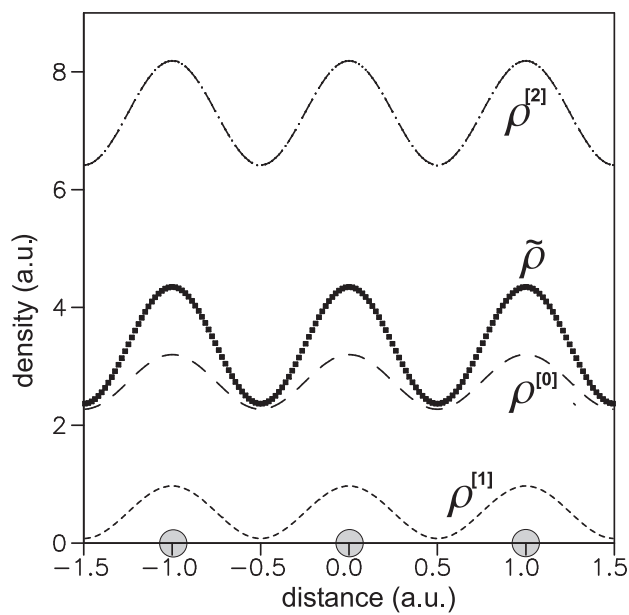}\end{center}

\caption{The electron densities for $\alpha=$2 a.u. Other
notations are the same as in Fig. \ref{fig:alpha_4}. Note that the
correct density (dots) is rather large between the localisation
centres.\label{fig:alpha_2}}
\end{figure}

Finally, we show in Fig. \ref{fig:alpha_2} the densities
calculated using both methods for orbitals which are least
localised when $\alpha_{1}^{*}>\alpha>\alpha_{2}^{*}$. The
density, obtained using the Fourier transform method, Eq.
(\ref{eq:final-density}), is spread almost uniformly in the
crystal volume and thus may correspond to a metallic band. At the
same time, the L\"owdin expansion method, Eq.
(\ref{eq:final-n-order-density}), does not converge at all and
the density is clearly diverges. %

One can expect that the latter situation can happen only for
metallic systems. Interestingly, our calculations (not reported
here) for such a realistic covalent system as a crystalline Si
show that the L\"owdin approach also fails in some cases when the
orbitals are not sufficiently localised. Note that various degree
of localisation of the orbitals can be obtained using different
localisation techniques and different choice of regions, see
\cite{Danyliv-LK-2004} for more details.

\section{Conclusions}

In summary, we have considered two numerical methods which allow
calculation of the electron density of a 3D periodic system
constructed via a set of non-orthogonal localised orbitals. The
first, so-called L\"owdin, method based on the power expansion of
the inverse of the overlap matrix has been found to be efficient
only for strongly localised orbitals. For an intermediate degree
of orbitals localisation this method has been found to be quite
computationally demanding since many terms in the series are to be
retained. However, if orbitals are not sufficiently localised (the
exact criterion has also been suggested), the method fails
altogether and the power expansion has been shown to be divergent.

Then, we have suggested another method based on the Fourier
transform of the localised orbitals which involves calculations of
inverse of only finite matrices and a $\mathbf{k}$ point summation
over the Brillouin zone. This method is computationally much less
demanding and does not have any convergency problems. Using a
simple model for the crystal electron density represented via a
set of Gaussian $s$ type orbitals in a simple cubic lattice (one
orbital per unit cell), we have shown that our method works
equally well within a rather wide range of orbitals having
different localisation, whereas the first method fails for a
relatively weakly localised orbitals. The application of the
Fourier transform method to realistic systems such as MgO and Si
perfect crystals is published elsewhere \cite{Danyliv-LK-2004}.

\section*{Acknowledgements}

We are extremely grateful to I. V. Abarenkov and I. Tupitsin for
useful and stimulating discussions. O.D. would also like to
acknowledge the financial support from the Leverhulme Trust (grant
F/07134/S) which made this work possible.

\end{document}